\documentclass{aastex631}

\usepackage{amsmath}
\usepackage{xcolor}
\def\r{\textcolor{red}}
\def\b{\textcolor{blue}}

\begin{document}

\title{The particle acceleration study in blazar jet}

\correspondingauthor{Hubing Xiao, Junhui Fan}
\email{hubing.xiao@shnu.edu.cn, fjh@gzhu.edu.cn}

\author[0000-0001-8244-1229]{Hubing Xiao}
\affiliation{Shanghai Key Lab for Astrophysics, Shanghai Normal University, Shanghai, 200234, China}

\author{Wenxin Yang}
\affiliation{Center for Astrophysics, Guangzhou University, Guangzhou 510006, China}
\affiliation{Key Laboratory for Astronomical Observation and Technology of Guangzhou, Guangzhou, 510006, China}
\affiliation{Astronomy Science and Technology Research Laboratory of Department of Education of Guangdong Province, \\
Guangzhou, 510006, China}

\author{Yutao Zhang}
\affiliation{Center for Astrophysics, Guangzhou University, Guangzhou 510006, China}

\author{Shaohua Zhang}
\affiliation{Shanghai Key Lab for Astrophysics, Shanghai Normal University, Shanghai, 200234, China}

\author[0000-0002-5929-0968]{Junhui Fan}
\affiliation{Center for Astrophysics, Guangzhou University, Guangzhou 510006, China}
\affiliation{Key Laboratory for Astronomical Observation and Technology of Guangzhou, Guangzhou, 510006, China}
\affiliation{Astronomy Science and Technology Research Laboratory of Department of Education of Guangdong Province, \\
Guangzhou, 510006, China}

\author{Liping Fu}
\affiliation{Shanghai Key Lab for Astrophysics, Shanghai Normal University, Shanghai, 200234, China}

\author{Jianghe Yang}
\affiliation{Department of Physics and Electronics Science, Hunan University
of Arts and Science, Changde 415000, China}
\affiliation{Center for Astrophysics, Guangzhou University, Guangzhou 510006, China}



\begin{abstract}
The particle acceleration of blazar jets is crucial to high-energy astrophysics, yet the acceleration mechanism division in blazar subclasses and the underlying nature of these mechanisms remain elusive.
In this work, we utilized the synchrotron spectral information (synchrotron peak frequency, $\log \nu_{\rm sy}$, and corresponding curvature, $b_{\rm sy}$) of 2705 blazars from the literature and studied the subject of particle acceleration in blazar jets by analysing the correlation between $\log \nu_{\rm sy}$ and $1/b_{\rm sy}$.
Our results suggested that the entire sample follows an energy-dependent probability acceleration (EDPA). 
Specifically, the low inverse Compton peak sources (LCPs) follow the mechanism that fluctuations of fractional gain acceleration (FFGA), while the high inverse Compton peak sources (HCPs) follow an acceleration mechanism of EDPA.
Our results indicated that the separation between LCPs and HCPs results from the electron peak Lorentz factor ($\gamma_{\rm p}$), and the differentiation should originate from different acceleration mechanisms.
Moreover, our study revealed a transition in the acceleration mechanism from FFGA to EDPA around $\log \nu_{\rm sy} \sim 15$ through a detailed analysis of binned-$\log \nu_{\rm sy}$. 
The mechanism of FFGA dominates the particle acceleration in LCP jets because of stronger jets and the EDPA dominates the particle energy gain in the HCPs due to a more efficient acceleration process.
\end{abstract}



\section{Introduction} \label{sec:intro}
Referring to the most distinctive and potent Active Galactic Nuclei (AGNs), blazars manifest exceptional observational properties. 
Notably, blazars exhibit multifaceted rapid variability across the entire wavelength bands, conspicuous high and variable polarization, powerful $\gamma$-ray emissions, apparent superluminal motion, and a confirmed core-dominated emission profile, as substantiated by a plethora of works \citep{Wills1992, Villata2006, Fan2002, Fan2014, Fan2021, Gupta2016, Xiao2019, Xiao2022MNRAS, Abdollahi2020, Pei2020RAA}.
These distinctive properties are ascribed to the presence of a powerful relativistic jet, which is pointing towards the observer, ensconced within the central supermassive black hole. 
Blazars are categorized into two distinct subclasses: BL Lacertae objects (BL Lacs) and flat spectrum radio quasars (FSRQs). 
Their emissions span from the radio to the very high energy (VHE) bands, exhibiting a characteristic bimodal structure in the spectral energy distribution (SED).

The broadband SED can be described by the radiation through relativistic particles interacting with magnetic fields, soft photons, or self-interactions and generating secondary particles and cascade process, namely, the leptonic model \citep{Blandford1979, Sikora1994, Blazejowski2000, Abramowski2015, Xue2019ApJ_1} and the hadronic model \citep{Mucke2001, Dimitrakoudis2012, Diltz2015}.
A common question in both blazar radiation mechanisms revolves around understanding the particle acceleration mechanism---how do these relativistic particles grow energy?
The observation that the radio-ultraviolet spectrum of blazars can be well-described by a log-parabolic function, as found by \citet{Landau1986}, is suggested to be linked to the particle acceleration mechanism, such as first-order Fermi acceleration. 
Subsequent studies extended this observation to the X-ray spectrum of blazars like Mrk 421 \citep{Massaro2004a} and Mrk 501, indicating a (quasi-)log-parabolic electron energy distribution (EED) \citep{Massaro2004b}. 
\citet{Massaro2004a} considered a simple statistical mechanism of particle energy gain, which could occur in a shock wave or a strong perturbation moving down a jet.
If the probability that a particle undergoes an acceleration step $i$, in which it has an energy gain independent of energy, then a power-law particle energy spectrum follows, otherwise, an energy-dependent probability acceleration (EDPA) forms a log-parabolic particle energy spectrum.
Instead of the acceleration probability being dependent on the particle energy, \citet{Tramacere2011} addressed the mechanism that fluctuations of fractional gain acceleration (FFGA) can form a log-parabolic particle energy spectrum.
In the framework of stochastic acceleration, \citet{Kardashev1962SvA} utilized the Fokker–Planck equation in the momentum-diffusion term and suggested that a `quasi-'monoenergetic particle injection can lead to a log-parabolic particle energy spectrum.

The anti-correlation between the synchrotron peak frequency $\nu_{\rm sy}$ and the curvature $b_{\rm sy}$ has been identified \citep{Massaro2006}, and exploring the correlation between these two parameters can provide insights into the particle acceleration processes in blazar jets \citep{Tramacere2007A&A521, Paggi2009A&A}.
\citet{Chen2014} summarized the correlation between $\log \nu_{\rm sy}$ and $1/b_{\rm sy}$ in different particle acceleration scenarios: 
(i) $\log \nu_{\rm sy} \approx 2/(5b_{\rm sy}) + C$ for the case of EDPA;
(ii) $\log \nu_{\rm sy} \approx 3/(10b_{\rm sy}) + C$ for the case of FFGA;
(iii) $\log \nu_{\rm sy} \approx 1/(2b_{\rm sy}) + C$ for the case of `quasi'-monoenergetic particle injection
in the framework of stochastic acceleration.
\citet{Chen2014} further studied the correlation between $\log \nu_{\rm sy}$ and $1/b_{\rm sy}$ and found a slope of 2.04. 
Their result suggested that the stochastic acceleration mechanism dominates the particle acceleration in blazar jets.
In a study by \citet{Tan2020}, a sample of 60 FSRQs yielded $1/b_{\rm sy} \propto 3.22 \log \nu_{\rm sy}$, consistent with the statistical acceleration of FFGA.
Similarly, \citet{Chen2023ApJ944} used a sample of 798 sources (504 FSRQs, 277 BL Lacs, 17 $\gamma$NLS1s), and obtained $1/b_{\rm sy} \propto 2.63 \log \nu_{\rm sy}$ for all blazars in their sample, $1/b_{\rm sy} \propto 2.55 \log \nu_{\rm sy}$ for FSRQs, and $1/b_{\rm sy} \propto 2.96 \log \nu_{\rm sy}$ for BL Lacs.
Meanwhile, \citet{Yang2022ApJS} used a larger sample of 2705 blazars and found $1/b_{\rm sy} \propto 1.97 \log \nu_{\rm sy}$ for blazars, $1/b_{\rm sy} \propto 2.86 \log \nu_{\rm sy}$ for FSRQs, and $1/b_{\rm sy} \propto 2.14 \log \nu_{\rm sy}$ for BL Lacs.

The reason for causing discrepancies among previous results regarding the particle acceleration mechanism of blazars and their subclasses may be twofold.
Firstly, the sample size could be insufficient for a comprehensive study of this subject. 
Secondly, there might be an evolution of particle acceleration mechanisms among blazars, contributing to the observed variations in the correlation.

In this work, following the previous works and taking advantage of the latest and largest SED fitting sample from \citet{Yang2022ApJS}, we aim to statistically study the particle acceleration mechanism and further investigate its evolution in blazars.
This paper is structured as follows:
In section \ref{sec:sample} we present our sample;
we present analysis and results in Section \ref{sec:result};
the discussions and conclusions will be given in Section \ref{sec:discussion};

\section{Sample}\label{sec:sample}
In order to study the particle acceleration mechanism and its possible evolution in blazars, we collected a sample of \textit{Fermi} blazars with available synchrotron peak frequency ($\log \nu_{\rm sy}$) and the corresponding curvature ($b_{\rm sy}$) from \citet{Yang2022ApJS}.
In total, we have 2705 sources, including 759 FSRQs, 1141 BL Lacs, and 808 blazar candidates of uncertain type (BCUs).

\section{Results}\label{sec:result}
\begin{figure}
\centering
\includegraphics[scale=1]{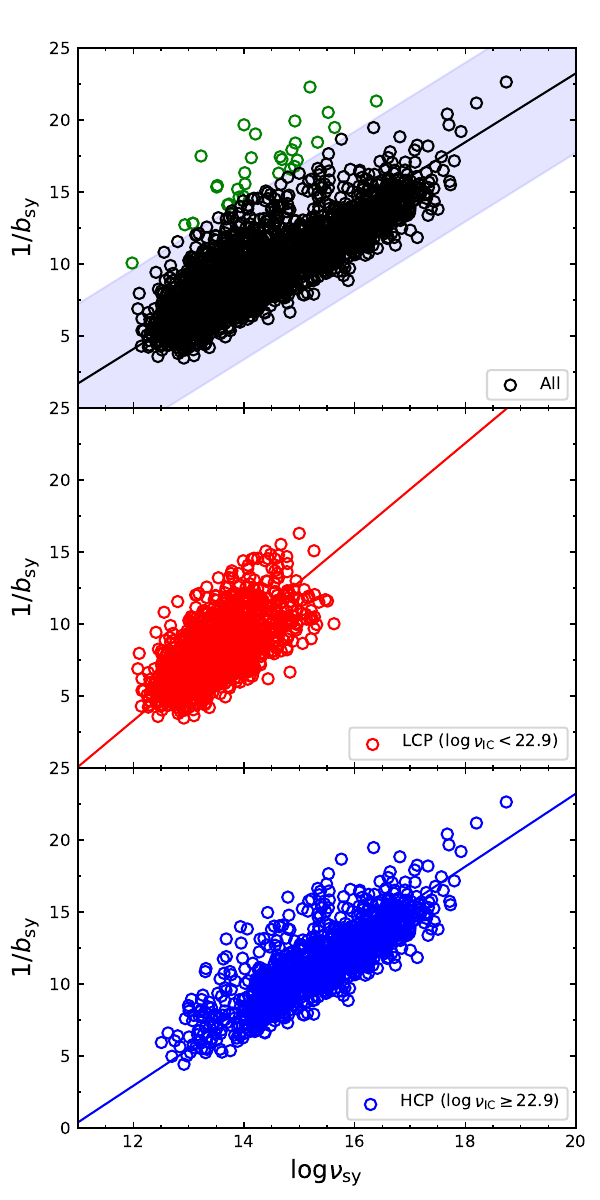}
\caption{The correlation between $1/b_{\rm sy}$ and $\log \nu_{\rm sy}$.
The correlation studies are investigated for all sources in our sample (in the upper panel), the light-blue shadow show a $[y- 3*{\rm std\_dev}, y+ 3*{\rm std\_dev}]$ interval and the green circles outside the shadow are scatters.
LCPs without scatters are shown in the middle panel, HCPs without scatters are shown in the bottom panel. 
The solid lines represent the best regression results.}
\label{b}
\end{figure}

As mentioned above, the acceleration mechanism can be tested by investigating the slope of the correlation between the peak frequency and the curvature for a sample of blazars. 
When the ordinary least square (OLS) bisector linear regression is adopted to the  present compilation, a correlation between $1/b_{\rm sy}$ and $\log \nu_{\rm sy}$, 
\begin{equation*}
1/b_{\rm sy} = (2.39 \pm 0.03)\log \nu_{\rm sy} - (24.51 \pm 0.39),
\end{equation*}
was obtained with a correlation coefficient $r=0.81$ and chance probability $p \sim 0$ through Pearson analysis for the entire sample as shown in the upper panel of Figure \ref{b}. 
We made excluded sources outside the light-blue shadow, which is defined by $[y- 3*{\rm std\_dev}, y+ 3*{\rm std\_dev}]$, and obtained a `clean' sample.
Meanwhile, the `clean' sample was divided into low IC peak frequency (LCP, $\log \nu_{\rm IC} < 22.9$) blazars and high IC peak frequency (HCP, $\log \nu_{\rm IC} \geq 22.9$) blazar based on \citet{Yang2023SCPMA}, the correlation between $1/b_{\rm sy}$ and $\log \nu_{\rm sy}$ for the LCPs and the HCPs are shown in the lower panel of Figure \ref{b}, the regression results
\begin{equation*}
1/b_{\rm sy} = (3.21 \pm 0.08)\log \nu_{\rm sy} - (35.24 \pm 1.02),
\end{equation*}
with $r=0.65$ and chance probability $p \sim 0$ for the LCPs;
\begin{equation*}
1/b_{\rm sy} = (2.54 \pm 0.04)\log \nu_{\rm sy} - (27.51 \pm 0.66),
\end{equation*}
with $r=0.81$ and chance probability $p \sim 0$ for the HCPs.

\begin{figure}
\centering
\includegraphics[scale=0.80]{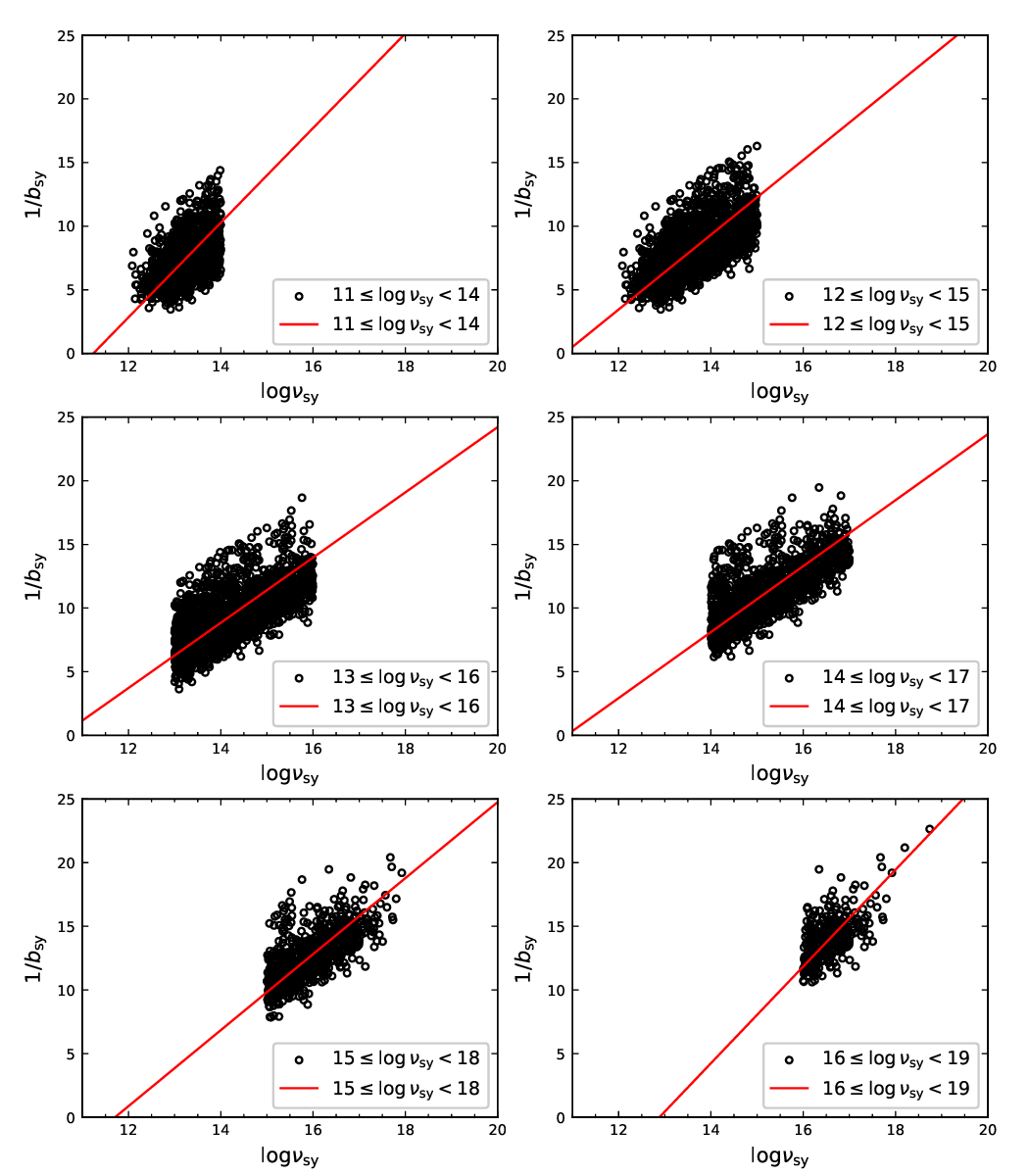}
\caption{The correlation between $1/b_{\rm sy}$ and $\log \nu_{\rm sy}$.
The red solid line represents the best OLS linear regression.}
\label{bin_fig}
\end{figure}

To investigate the possible evolution of the particle acceleration mechanism in blazars, we assume $1/b_{\rm sy} = \beta \log \nu_{\rm sy} + \alpha$.
Subsequently, we calculate $\beta$ for sources falling within each bin ranging from $\log \nu_{\rm sy}$ to $\log \nu_{\rm sy} + 3$, $\log \nu_{\rm sy}$ is increased incrementally by one to establish distinct bins for computing $\beta$. 
The corresponding statistical results are shown in Figure \ref{bin_fig} and detailed in Table \ref{bin_tab}.

\begin{table}[htbp]
\centering
\caption{The evolution of $\beta$ for sources with increasing $\log \nu_{\rm syn}$.}
\label{bin_tab}
\begin{tabular}{lcccccc}
\hline
Bin ID & Energy range & N & $\beta$ & $\alpha$ & r & p  \\
(1) & (2) & (3) & (4) & (5) & (6) & (7) \\
\hline
1 & $11 \leq \log \nu_{\rm sy} < 14$	& 1318	& $3.72 \pm 0.12$	& $-41.78 \pm 1.64$	  & 0.50 & $9.0 \times 10^{-85}$   \\
2 & $12 \leq \log \nu_{\rm sy} < 15$	& 1903	& $2.94 \pm 0.06$	& $-31.79 \pm 0.76$	  & 0.64 & $3.2 \times 10^{-224}$  \\
3 & $13 \leq \log \nu_{\rm sy} < 16$	& 1970	& $2.56 \pm 0.04$	& $-27.03 \pm 0.58$	  & 0.70 & $3.0 \times 10^{-290}$  \\
4 & $14 \leq \log \nu_{\rm sy} < 17$	& 1313	& $2.59 \pm 0.05$	& $-28.17 \pm 0.70$	  & 0.70 & $1.3 \times 10^{-197}$  \\
5 & $15 \leq \log \nu_{\rm sy} < 18$	& 769	& $2.99 \pm 0.08$	& $-34.98 \pm 1.25$	  & 0.67 & $6.6 \times 10^{-102}$  \\
6 & $16 \leq \log \nu_{\rm sy} < 19$	& 343	& $3.80 \pm 0.19$	& $-48.97 \pm 3.08$	  & 0.61 & $6.7 \times 10^{-37}$  \\
\hline
\end{tabular}
\tablecomments{
column (1) gives the range ID;
column (2) gives the range of taking sources;
column (3) is the number of the sources in the corresponding range;
column (4) gives the $\beta$;
column (5) gives the $\alpha$;
column (6) gives the Pearson correlation coefficient;
column (7) gives the chance probability;
}
\end{table}

\section{Discussion}\label{sec:discussion}
\subsection{The physical meaning of dividing boundary with $\log \nu_{\rm IC}$}

In the frame of a leptonic model, the synchrotron peak frequency is written as
\begin{equation}
    \nu_{\rm sy} = 3.7 \times 10^{6}\, \gamma_{\rm p}^{2} \, B\, \frac{\delta}{1+z} \, {\rm Hz},
\label{nu_sy}
\end{equation}
where $B$ is the magnetic field strength in units of Gauss \citep{Tavecchio1998}, $\gamma_{\rm p}$ is the Lorentz factor of electrons that contribute most to the synchrotron peak and $\delta$ is a Doppler beaming factor.
In the Thomson regime, the peak frequency of the IC peak frequency is given by 
\begin{equation}
    \nu_{\rm SSC} = \frac{4}{3}\, \gamma_{\rm p}^{2} \, \nu_{\rm sy},
\label{nu_ssc}
\end{equation}
in the case of the SSC model; while in the case of the EC process, soft photons are fed externally, and the peak frequency is given by
\begin{equation}
    \nu_{\rm EC} = \frac{4}{3}\, \gamma_{\rm p}^{2} \, \nu_{\rm ext} \, \frac{\Gamma \delta}{1+z},
\label{nu_ec}
\end{equation}
where, $\nu_{\rm ext}$ is the frequency of external photons, $\nu_{\rm ext} = 2.46 \times 10^{15} \, {\rm Hz}$ for the case of external photons coming from the BLR and $\nu_{\rm ext} = 7.7 \times 10^{13} \, {\rm Hz}$ for the case of external photons coming from the DT \citep{Tavecchio2008MNRAS386, Ghisellini2015MNRAS}, and $\Gamma$ is a bulk Lorentz factor.

\citet{Yang2023SCPMA} estimated the IC peak frequency ($\nu_{\rm IC}$) through fitting the IC bump with a parabolic function, studied the distribution of $\log \nu_{\rm IC}$ and found bimodal structure via the Bayesian method.
They suggested using $\log \nu_{\rm IC} = 22.9$ to separate blazars into two classes, namely the low IC peak frequency ($\log \nu_{\rm IC} < 22.9$, LCP) blazars and high IC peak frequency ($\log \nu_{\rm IC} \geq 22.9$, HCP) blazars.
While the nature of separating the LCP and the HCPs remaining unclear.
The extragalactic background absorption (EBL), in the high and very high energy $\gamma$-ray bands, could affect the location of IC peak, the neglection or underestimation of EBL could bias the $\nu_{\rm IC}$ and consequent affect the separation value.

\begin{figure}
\centering
\includegraphics[scale=1]{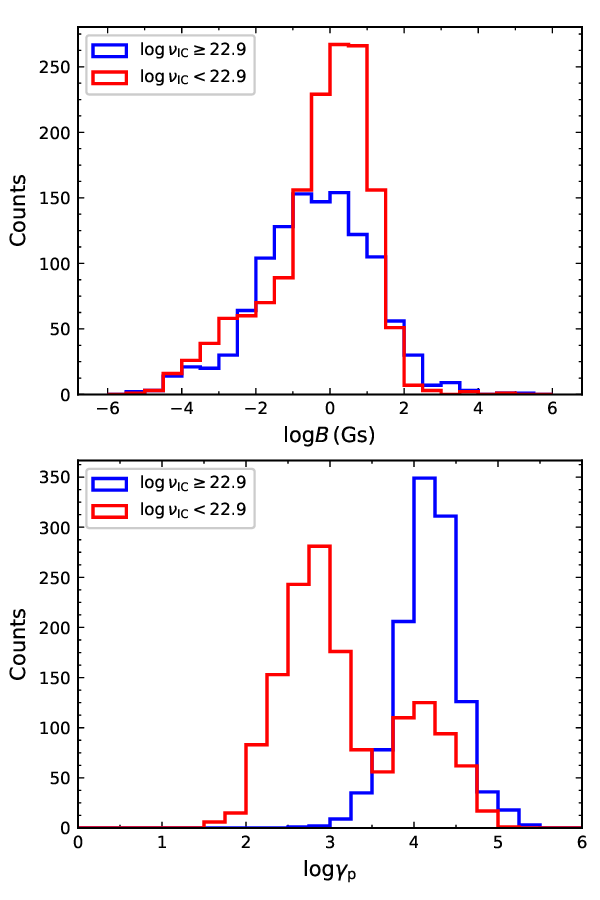}
\caption{The correlation between $1/b_{\rm sy}$ and $\log \nu_{\rm sy}$.
The red histogram stands for the sources with $\log \nu_{\rm IC} < 22.9$, and the blue histogram stands for the sources with $\log \nu_{\rm IC} \geq 22.9$.}
\label{dis}
\end{figure}

The LCPs distributed tighter than the HCPs in the space of $\log \nu_{\rm sy}$, and the HCPs should have a averagely larger $\log \nu_{\rm sy}$ than the LCPs, seen in the middle and bottom panels of Figure \ref{b}.
Equation (\ref{nu_sy}) indicates that the synchrotron peak frequency primarily depends on $\delta$, $B$, and $\gamma_{\rm p}$, and a comprehensively joint effort of these three parameters results in the divergence in $\log \nu_{\rm sy}$.
The Doppler beaming factor, $\delta$, has been consistently estimated in various studies, yielding values around $\sim 10{\rm s}$ for the majority of blazars \citep[e.g.,][]{Liodakis2018, Zhang2020, Pei2022ApJ, Xiao2022ApJ_1}. 
Consequently, a dispersion in $\delta$ is not expected to be a main reason of demonstrating difference in $\log \nu_{\rm sy}$.
In our recent work, \citet{Fan2023ApJS} estimated $B$ and $\gamma_{\rm p}$ for blazars with certain classes and they provided lower and upper limits of $B$ and $\gamma_{\rm p}$ for those BCUs by treating these sources as FSRQs and BL Lacs, respectively.
We obtained the values of $B$ and $\gamma_{\rm p}$ for BCUs using the predicted class assigned by the Fractal Dimension - Inverse Discrete Wavelet Transform (FDIDWT) method introduced in \citet{Cao2024ApJ}.
Figure \ref{dis} illustrates the distribution of $B$ and $\gamma_{\rm p}$ for the HCPs and the LCPs.
The analysis reveals that the LCPs have $\log B = -0.39\pm 1.50$, while the HCPs have $\log B = -0.29\pm 1.39$.
A student t-test yields $p=0.08$, indicating that the hypothesis of these two samples having identical average values cannot be rejected.
Therefore, $B$ is also not the main reason separating the LCPs and the HCPs in $\log \nu_{\rm sy}$.
Then, the focus shifts to $\gamma_{\rm p}$, and as depicted in the lower panel of Figure \ref{dis}, it is evident the HCPs and parts of the LCPs have averagely larger $\log \gamma_{\rm p}$ than the majority of the LCPs.
 
Based on equations (\ref{nu_ssc}) and (\ref{nu_ec}), this differentiation in $\log \gamma_{\rm p}$ directly translates to the divergence in IC peak frequency $\log \nu_{\rm IC}$ between HCPs and LCPs.
In case of SSC process, it is evident that the value of $\nu_{\rm SSC}$ is mainly depends on $\gamma_{\rm p}$ because, as we have mentioned above, $\nu_{\rm sy}$ is mainly affected by $\gamma_{\rm p}$.
In the case of EC process, the cooling process from seed photons, namely $\nu_{\rm ext}$, also play an important role in determining $\nu_{\rm EC}$. 
The cooling effect has been counted in the estimation of $B$ and $\gamma_{\rm p}$ in \citet{Fan2023ApJS}, where we obtained the values of $B$ and $\gamma_{\rm p}$ for the discussions presented in this work.
To sum up, the separation in $\log \nu_{\rm IC}$ between LCP and HCP mainly stems from the dispersion in $\gamma_{\rm p}$.

\subsection{The particle acceleration in blazar jets}
Based on the sample, we studied the correlation between $1/b_{\rm sy}$ and $\log \nu_{\rm sy}$.
Our findings reveal a relationship characterized by $1/b_{\rm sy} \propto 2.39\log \nu_{\rm sy}$, indicative of particle acceleration in blazar jets occurring with EDPA. 
Furthermore, this correlation implies a trend towards smaller curvatures as the synchrotron peak frequency increases.
This is attributed to the shorter cooling timescale compared to the timescales for acceleration \citep{Tramacere2009}.

There are 31 scatters, with relatively large $1/b_{\rm sy}$ in the entire sample, are marked in green in the upper-left region of the upper panel of Figure \ref{b}.
These scatters are caused by the lack of X-ray data, result in a consequent insufficient constrain on the fitting of synchrotron bump and a small value of curvature $b_{\rm sy}$.
In the middle and bottom panels of Figure \ref{b}, the sources are divided with the boundary of $\log \nu_{\rm IC} = 22.9$.
The value of the boundary is intricately linked to the $B$, $\delta$ and $\gamma_{\rm p}$ and is mainly determined by $\gamma_{\rm p}$, we found that the particles in LCP jets are accelerated with FFGA and in HCP jets are accelerated in the form of EDPA.
Moreover, the acceleration mechanism of EDPA is more efficient than FFGA, because the HCPs have an averagely larger $\log \gamma_{\rm p}$ than the LCPs as shown in the middle and bottom panels of Figure \ref{b}. 

If the separation between LCPs and HCPs is caused by the difference in the electron Lorentz factor, resulting from distinct particle acceleration mechanisms. 
The transition, which should not be accomplished suddenly but gradually, from one mechanism to another is expected to leave tracks. 

\begin{figure}
\centering
\includegraphics[scale=0.8]{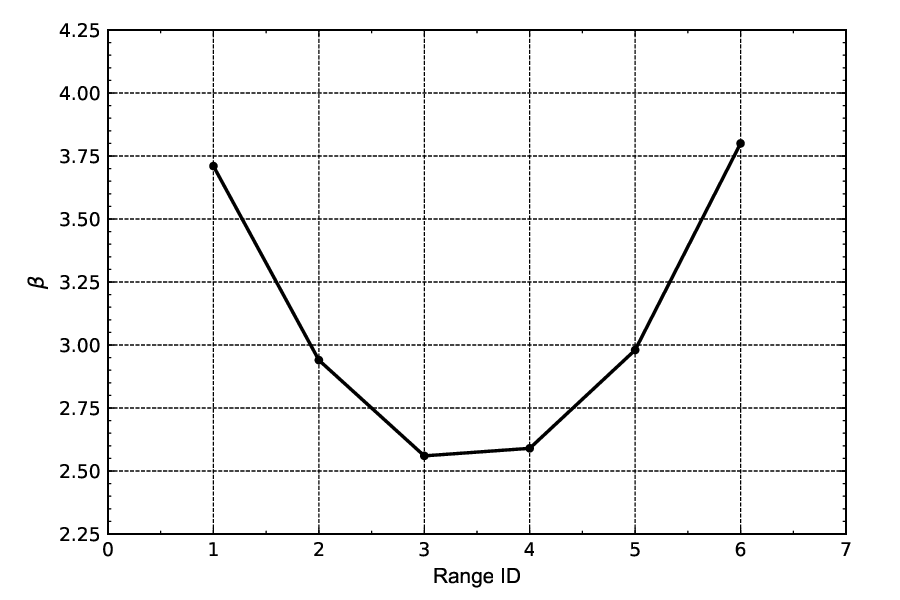}
\caption{The slope of correlation $1/b_{\rm sy}$ against $\log \nu_{\rm sy}$ in range of $[\log \nu_{\rm sy},  \log \nu_{\rm sy} +3)$.}
\label{slope}
\end{figure}

To explore the possible evolution or transition between these acceleration mechanisms, we divided the sample into six bins for $\log \nu_{\rm sy}$ with IDs ranging from 1 to 6.
We then examined the slope ($\beta$) for each bin, with the results presented in Table \ref{bin_tab} and illustrated in Figure \ref{bin_fig}. 
The change of $\beta$ is illustrated in Figure \ref{slope}, it is clear that $\beta$ decreases from 3.71 to 2.56 and subsequently increases to 3.80.
The values of 3.71 and 3.80, corresponding to `Bin 1' and `Bin 6', should be taken very carefully because these two bins contain borders and the quantity of the sources near the bin border may significantly influence the slope.
In general, the FFGA mechanism seems to dominate the sources with relatively small $\log \nu_{\rm sy}$, then the EDPA mechanism starts to dominate when the $\log \nu_{\rm sy}$ gets larger.

The first acceleration mechanism transition from FFGA to EDPA in Figure \ref{slope}, namely the slope change from `Bin 2' to `Bin 3' or `Bin 4', and the transition likely occurs at $\log \nu_{\rm sy} \sim 15$.
Interestingly, most of the LCPs occupy the region $\log \nu_{\rm sy} \lesssim 15$ and the HCPs occupy the region $\log \nu_{\rm sy} \gtrsim 15$.
Thus, the first acceleration mechanism transition in Figure \ref{slope} is consistent with the acceleration mechanism transition from the LCPs to the HCPs, we suggest the first transition in Figure \ref{slope} is interpreted as the separation of $\log \gamma_{\rm p}$.

However, the second transition from EDPA to FFGA in Figure \ref{slope}, namely the slope change from `Bin 3' or `Bin 4' to `Bin 5' or `Bin 6', maybe caused by those sources near the bin border and should be carefully taken as a real transition, as we can see in the bottom-left panel of Figure \ref{bin_fig}. 

\subsection{The acceleration mechanism}
According to our results, the statistical acceleration mechanism dominates the energy gain of particles in the blazar jets and the mechanism transition is derived.
In the framework of statistical acceleration, the EDPA mechanism was discussed in \citet{Massaro2004a}.
In this mechanism, the accelerated particle's energy is expressed as $\gamma_{i} = \varepsilon \gamma_{i-1}$, in which the energy gain $\varepsilon$ of acceleration step $i$ is independent of energy and assumed to be constant.
However, the probability of the particle undergoes acceleration step $i$ is dependent on the particle energy as $p_{i} = g/\gamma_{i}^{q}$, where $g$ and $q$ are positive constants.
Thus, a particle with a initial Lorentz factor $\gamma_{0}$ undergoing $n$ step of acceleration has a probability of 
\begin{equation}
\Pi^{n}_{i=0} p_{i} = g^{n}/ \Pi^{n}_{i=0} \gamma_{i}^q = g^{n} / [\gamma_{0}^{nq} \varepsilon^{\frac{n(n-1)}{2} q}].
\label{prob}
\end{equation}
This leads to the final accelerated energy
\begin{equation}
\gamma_{n} = \varepsilon^{n} \gamma_{0}.
\label{g1}
\end{equation}

An alternative mechanism known as FFGA was introduced by \citet{Tramacere2011}. 
In the FFGA model, the energy gain is expressed as $\gamma_{i} = \varepsilon_{i} \gamma_{i-1}$, where $\varepsilon_{i} = \bar{\varepsilon} + \chi_{i}$. 
Here, $\bar{\varepsilon}$ represents the systematic energy gain, and $\chi_{i}$ is a random variable with a probability density function having a zero mean value ($\langle \chi \rangle = 0$) and a variance of $\delta_{\chi}$.
For a particle with an initial Lorentz factor $\gamma_{0}$ undergoing $n$ steps of acceleration in the FFGA model, the expected final Lorentz factor is given by:
\begin{equation}
\gamma_{n} = \gamma_{0} \Pi_{i=0}^{n} \varepsilon_{i} = \gamma_{0} \Pi_{i=0}^{n} (\bar{\varepsilon} + \chi_{i}) = \gamma_{0} \bar{\varepsilon}^{n} \Pi^{n}_{i=0}{(1 + \chi_{i}/\bar{\varepsilon})}.
\label{g2}
\end{equation}

Taking into account that both EDPA and FFGA mechanisms are simultaneously at play in blazar jets, the dominance of one over the other depends on the specific blazar subclass. 
For LCPs, FFGA appears to be the dominant mechanism, while EDPA is more prominent in HCPs.
In both acceleration mechanisms, the accelerated particle's energy is influenced by the number of acceleration steps ($n$). 
This parameter serves as a crucial variable determining the final Lorentz factor ($\gamma_{n}$) in both scenarios, as described by equations \ref{g1} and \ref{g2}.

For a given synchrotron intensity with corresponding electron Lorentz factor $\gamma_{\rm p}$, the formation of relativistic particles through EDPA involves two possible approaches:
(1) according to equation \ref{prob}, a substantial number of particle supplements is provided due to the decreasing acceleration probability with each acceleration step $i$.
In this scenario, one should anticipate a greater supply of material to provide more seed electrons, and thus those sources with a higher accretion ratio (e.g., FSRQ, \citealp{Ghisellini2011, Sbarrato2012, Xiao2022ApJ_1, Pei2022ApJ}) should have this privilege;
(2) alternatively, with a relatively large energy gain $\varepsilon$ in each step of acceleration, the acceleration process becomes more efficient, and particles can attain $\gamma_{\rm p}$ in fewer steps before the acceleration probability diminishes.
We found that the particle acceleration in HCP jets is dominated by EDPA, while the HCPs primarily consist of BL Lacs, which are believed to have a smaller accretion ratio.
Thus, the acceleration may be more efficient in the jets of HCPs and particles gain energy significantly in each acceleration process.  

Similarly, for a given synchrotron intensity with corresponding electron Lorentz factor $\gamma_{\rm p}$.
According to the equation \ref{g2}, the particles can be accelerated to $\gamma_{\rm p}$ efficiently when the $\chi_{i}$ is significantly positive more often than negative in the case of FFGA.
This condition indicates a relatively strong acceleration process.
In this work, we noticed that the particle acceleration in LCP jets is dominated by FFGA, and the 1500 LCPs in this work (consisting of 705 FSRQs, 538 BCUs and 257 BLLs) are mainly constructed by FSRQs.
This preference for a relatively strong acceleration process aligns with the understanding that FSRQs generally exhibit higher jet power compared to BL Lacs \citep{Chen2023ApJS268}.

\section{Conclusion}
In this work, for the purpose of studying the particle acceleration mechanism in blazar jets, we conducted an analysis of the correlation between $1/b_{\rm sy}$ and $\log \nu_{\rm sy}$ for the entire sample and its subgroups (LCP and HCP).
Our findings can be summarized as follows:
(1) A slope of 2.39 indicates an EDPA for the entire sample, a slope of 3.21 indicates a FFGA for the LCP, and a slope of 2.54 indicates an EDPA for the HCP;
(2) The separation between LCPs and HCPs is attributed to the distinct $\gamma_{\rm p}$ values between the two subclasses;
(3) Further exploration of the particle acceleration mechanism involved an investigation of the slope changes in different $\log \nu_{\rm sy}$ bins. 
This analysis revealed an acceleration mechanism transition from FFGA to EDPA around $\log \nu_{\rm sy} \sim 15$;
(4) A comparison between EDPA and FFGA suggested that EDPA is more efficient in the HCPs, where particles experience significant energy gain in each acceleration process.
Conversely, in LCPs, a stronger acceleration process operates through FFGA due to the presence of more powerful jets.

\begin{acknowledgments}
H.B.X. acknowledges the support from the National Natural Science Foundation of China (NSFC 12203034), the Shanghai Science and Technology Fund (22YF1431500), and the science research grants from the China Manned Space Project.
J.H.F acknowledges the support from the NSFC U2031201, NSFC 11733001, the Scientific and Technological Cooperation Projects (2020–2023) between the People’s Republic of China and the Republic of Bulgaria, the science research grants from the China Manned Space Project with No. CMS-CSST-2021-A06, and the support for Astrophysics Key Subjects of Guangdong Province and Guangzhou City.
This research was partially supported by the Bulgarian National Science Fund of the Ministry of Education and Science under grants KP-06-H38/4 (2019), KP-06-KITAJ/2 (2020) and KP-06-H68/4 (2022).
S.H.Z acknowledges support from the National Natural Science Foundation of China (Grant No. NSFC-12173026), the Program for Professor of Special Appointment (Eastern Scholar) at Shanghai Institutions of Higher Learning and the Shuguang Program of Shanghai Education Development Foundation and Shanghai Municipal Education Commission.
L.P.F acknowledges the support from the NSFC grants 11933002, STCSM grants 18590780100, 19590780100, SMEC Innovation Program 2019-01-07-00-02-E00032, and Shuguang Program 19SG41.
\end{acknowledgments}

\bibliography{lib}{}
\bibliographystyle{aasjournal}



\end{document}